\documentclass{cernphprep}
\usepackage{xcolor,pict2e,curve2e}
\begin{document}

\newcommand{\Wtaunu}{\ensuremath{W^{\pm}\to\tau^{\pm}\nu}}
\newcommand{\Htaunu}{\ensuremath{H\to\tau^{\pm}\nu}}
\newcommand{\WHtaunu}{\ensuremath{W/H\to\tau^{\pm}\nu}}
\newcommand{\ZHtautau}{\ensuremath{Z/H\to\tau^{-}\tau^{+}}}
\newcommand{\Ztautau}{\ensuremath{Z/\gamma^{*}\to\tau^{-}\tau^{+}}}
\newcommand{\Zee}{\ensuremath{Z/\gamma^{*}\to e^{-}e^{+}}}
\newcommand{\Zmm}{\ensuremath{Z/\gamma^{*}\to\mu^{-}\mu^{+}}}
\newcommand{\Htautau}{\ensuremath{H\to\tau^{-}\tau^{+}}}
\newcommand{\TauNu}{\ensuremath{\tau^{\pm}\nu}}
\newcommand{\TauTau}{\ensuremath{\tau^{-}\tau^{+}}}
\newcommand{\Tau}{\ensuremath{\tau}}
\newcommand{\TauMin}{\ensuremath{\tau^-}}
\newcommand{\TauPlus}{\ensuremath{\tau^+}}
\newcommand{\RHO}{\ensuremath{\pi^{-}\pi^{0}\nu}}
\newcommand{\ARES}{\ensuremath{\pi^{-}\pi^{0}\pi^{0}\nu}}
\newcommand{\ARESMULTI}{\ensuremath{\pi^{-}\pi^{-}\pi^{+}\nu}}
\newcommand{\PI}{\ensuremath{\pi^{-}\nu}}
\newcommand{\KA}{\ensuremath{K^{-}\nu}}
\newcommand{\PTAUZ}{\ensuremath{p_{\tau}^{\mathrm{Z}}}}
\newcommand{\PTAU}{\ensuremath{P_{\tau}}}
\newcommand{\ZGAMMA}{\ensuremath{Z/\gamma^{*}}}
\newcommand{\HIGGS}{\ensuremath{H}}
\newcommand{\CHHIGGS}{\ensuremath{H^{\pm}}}
\newcommand{\CHW}{\ensuremath{W^{\pm}}}

\begin{titlepage}
\PHnumber{2011--307}
\PHdate{December 2011}

\title{ {\tt TauSpinner} program for studies on spin effect in tau production at the LHC.
}

\begin{Authlist}
Z. Czyczula\Iref{a4}, T. Przedzinski\IAref{a2}{*}, Z.Was\IIref{a1}{a3}
\end{Authlist}

\Instfoot{a1}{Institute of Nuclear Physics, Cracow, Poland}
\Instfoot{a2}{Institute of Physics, Jagiellonian University, Cracow, Poland}
\Instfoot{a3}{CERN, CH-1211, Geneva 23, Switzerland, Theory Group, Physics Department}
\Instfoot{a4}{Yale University, New Haven, USA}
\Anotfoot{*}{The work of Tomasz Przedzinski was supported in part by the Polish Government grant NN202127937 (years 2009-2011).}


\begin{abstract}
Final states involving tau leptons are important components of
searches for new particles at the Large Hadron Collider (LHC). 
A proper treatment of tau spin
effects in the Monte Carlo (MC) simulations is important for
understanding the detector acceptance as well as for 
the measurements of tau polarization and tau spin correlations.
In this note we present a {\tt TauSpinner} package designed 
to simulate the spin effects. 
It relies on the availability of the four-momenta of the taus and their 
decay products in the analyzed data.
The flavor and the four-momentum of the boson decaying to the \TauTau~or
\TauNu~pair need to be known. In the \ZGAMMA~case the initial state quark configuration
is attributed from the intermediate boson
kinematics, and the parton distribution functions (PDF's).
{\tt TauSpinner} is the first algorithm suitable for emulation of tau spin effects
in tau-embedded samples. It is also the first tool that offers the user the flexibility 
to simulate a desired spin effect at the analysis level. An algorithm to attribute tau helicity states to a previously generated sample is also provided.
\end{abstract}
\Submitted{(Submitted to Eur. Phys. J. C )}
\end{titlepage}
\section{Introduction}

Tau leptons are an excellent signature with which to
probe new physics at the LHC. As the heaviest leptons, they
have the largest coupling to the Higgs boson both
in the Standard Model (SM) and the Minimal
Supersymmetric Standard Model (MSSM). Their short-enough lifetime and parity-violating decays
allow for their spin information to be preserved in the decay product 
kinematics recorded in the detector. They are therefore the only leptons suitable
for measuring longitudinal polarization and its correlations,
which provide important constraints on the nature of the observed resonance.

In this note we present a {\tt TauSpinner} package, which is a MC 
program designed to generate tau spin effects in {\it any} tau sample
provided their origin is known. This tool has two important applications:
\begin{description}
\item[Data driven analysis] of \Ztautau~and \Wtaunu~\\ backgrounds. 
In the case of the algorithms for embedding taus on measured light lepton samples~\cite{EMBEDDED} 
the {\tt TauSpinner} may represent a third step. The first step is to construct the tau four-momenta
from the four-momenta of the measured lighter leptons (and accompanying photons) while the second step
comprises the decay of unpolarized taus using\footnote{In order to enhance statistics for the spin analysis, the decay of unpolarized taus
can be performed multiple times.}\\ e.g. {\tt Tauola}~\cite{TAUOLA1,TAUOLA2,TAUOLA3}. Omission of the last step leaves the kinematics of the tau decay particle different from
what is expected from the polarized taus that appear in nature. 
This could lead to a mis-modeling of the shapes of various observables and hence 
a mis-measurement of the acceptance of a given set of cuts.
\item[MC studies.] 
Since the tau spin effects affect the overall acceptance of the taus 
in the detector, a proper treatment of these effects
is of great importance for the interpretation of the results as well as feasibility studies of new
models.
It is also key for measurements of tau polarization and tau spin correlations. 
The {\tt TauSpinner} algorithm allows one to create samples with different
tau polarizations from an initial sample by re-weighting events to give the desired distributions as a function of the decay mode of the tau. Furthermore, 
the helicity states of the taus can be attributed to the previously 
generated sample (with spin effect included or introduced with the {\tt TauSpinner}
weight).
\end{description}

This letter is organized as follows.
In Sec.~\ref{sect:TheAlg} we present the algorithm while
in Sec.~\ref{sect:Perf} we discuss its performance as compared to
the standard {\tt Tauola} MC package.
The results are summarized in Sec.~\ref{sect:Summary}. 
All relevant technical details are provided in App. A and B.

\section{The TauSpinner algorithm}~\label{sect:TheAlg}

The {\tt TauSpinner} algorithm relies on a leading order approximation in which 
spin amplitudes are used to calculate the spin density matrices for hard
$2\to2$~or $1\to2$ Born level processes
~\cite{TauSpinERWZW,Davidson:2010rw,Jadach:1984iy,Jadach:1993yv,Jadach:1999vf}. 
{\tt TauSpinner} is constrained to the longitudinal tau spin degrees only.
It starts with identifying the flavor of
the intermediate boson: \CHW, \ZGAMMA, \HIGGS~or \CHHIGGS. The information on the four-momenta of the
outgoing taus and their decay products as well as the intermediate
boson four-momentum is then used to determine the polarimetric vectors.
The longitudinal tau polarization (\PTAU) is randomly generated as specified in Tab.~\ref{tab:SpinConf} and set
to $\pm 1$ which correspond to pure tau helicity states.
Probability of the helicity states is a constant for taus originating from the \CHW, \CHHIGGS~and \HIGGS~bosons.
In \Ztautau~events, the probability, denoted \PTAUZ, is a function of the \TauMin~scattering angle, $\theta$, and
the center of mass squared of the hard process, $s$. At the Born
level, and in the ultrarelativistic limit,~\PTAUZ~is, following notations of~\cite{TauSpinERWZW}, given by:

\begin{equation}
\label{eq:1}
\PTAUZ(s,\theta)=\frac{\frac{\mathrm{d}\sigma}{\mathrm{d}\cos\theta}(s, \cos\theta,P_\tau=1)}{\frac{\mathrm{d}\sigma}{\mathrm{d}\cos\theta}(s, \cos\theta,P_\tau=1)+\frac{\mathrm{d}\sigma}{\mathrm{d}\cos\theta}(s, \cos\theta, P_\tau,=-1)}
\end{equation}

\noindent where

\begin{equation}
\label{eq:2}
\frac{\mathrm{d}\sigma}{\mathrm{d}\cos\theta}(s, \cos\theta,P_\tau)=(1+\cos^2\theta)F_\mathrm{0}(s)+2\cos\theta F_\mathrm{1}(s)\\ 
-P_\tau[(1+\cos^2\theta) F_\mathrm{2}(s)+2\cos\theta F_\mathrm{3}(s)], 
\end{equation}

\noindent and $F_\mathrm{i}(s)$'s are four form factors which depend on the initial and
the final state fermion couplings to the $Z$ boson and the propagator.
The dependence on the $\tau^-$ longitudinal polarization  $P_\tau$ is taken into account 
(note  that in this case $P_\tau=\pm$1, also  $P_{\tau_1}=P_{\tau_2}=P_\tau$).
In~\cite{TauSpinERWZW} the probability \PTAUZ~is calculated using the information on the initial state quarks stored
at the generation level.
In {\tt TauSpinner} the initial state quark configuration is attributed 
stochastically from the intermediate boson
kinematics and PDFs in the following steps:
\begin{enumerate}
\item
The invariant mass of the \ZGAMMA~is calculated from 
the intermediate boson four-vector. Note that it does not need to coincide
with the sum of \TauTau~four-momenta as it may include photons
of final state bremsstrahlung.
\item The scattering angle, $\cos\theta$, is calculated in the \TauTau \\
pair rest frame from the angle between the direction of the first beam (1,0,0,1)
boosted to this frame and the direction of the \TauPlus~or from the angle 
between the direction of the second beam (1,0,0,-1) boosted to this frame
and the direction of the \TauMin. In the final step the average of the two~\footnote{In ~\cite{Was:1989ce} this angle is referred to as $\theta^*$ is taken~\cite{Was:1989ce}}.
\item The fraction of momenta taken by partons of the proton: $x_1$ and $x_2$ are resolved from
constraints\\ $x_\mathrm{1}x_\mathrm{2} E_\mathrm{CM}^2 = s$ and $ (x_\mathrm{1} - x_\mathrm{2}) E_\mathrm{CM}={p}_\mathrm{z}$, where
$E_\mathrm{CM}$\\~and $p_z$~denote the collision center of mass energy and the longitudinal component of the \ZGAMMA~momentum, respectively.
\item Probabilistic choice on the basis on the leading order $2\to 2$ Born level cross sections
and PDF's is performed to attribute the flavors to the incoming
quarks and sign of $\theta$. 
\end{enumerate}

The probability \PTAUZ~is calculated as a weighted average over all possible initial state quark configurations. 
This solution is only implemented for the case of proton-proton ($pp$) collisions.
For other types of collision events
the functionality of {\tt TauSpinner} for \Ztautau process is restricted to 
longitudinal spin correlation only (\PTAUZ=0.5).

Note that the calculation of the momentum fractions $x_\mathrm{1}$~and $x_\mathrm{2}$~is performed within a collinear approximation where
the initial state radiation quarks and gluons are assumed 
to have no transverse momentum. Furthermore
the final state radiation photons are taken into account at collinear level 
only: they are omitted from the boson decay vertex. Any deviation 
from the energy-momentum conservation in the boson decay vertex is 
attributed to the presence of the photons.
As shown in Sec.~\ref{sect:Perf}
these approximations have practically no impact on the implemented spin effects.

\begin{table}
\begin{center}
 \caption{Probability for the configuration of the longitudinal
  polarization of taus from different origins~\cite{TauSpinERWZW}. }
 \label{tab:SpinConf}
\begin{small}
\begin{tabular}{lccc}
\hline\noalign{\smallskip}
Origin & $P_{\tau_1}$ & $P_{\tau_2}$ &Probability\\
\hline\noalign{\smallskip}
Neutral Higgs bosons: \HIGGS &
+1 & --~1&0.5\\
& --~1 & +1&0.5\\
Neutral vector boson: \ZGAMMA&
+1 & +1&\PTAUZ\\
& --~1 & --~1&1--\PTAUZ\\
Charged Higgs: \CHHIGGS &
+1 & --~ &1.0\\
Charged vector boson: \CHW&
--~1 & --~ &1.0\\
\hline\noalign{\smallskip}
  \end{tabular}
 \end{small}
 \end{center}
\end{table}

\subsection{Calculation of the spin weight}

The outcome of running the {\tt TauSpinner} program is 
a spin weight attributed to each event separately.

In the \TauNu~ final state, for any decay of a polarized tau, the spin weight is defined as~\cite{TAUOLA1,TAUOLA2,TAUOLA3}: 

\noindent
\begin{equation}\label{eq:3}
w_\mathrm{T}=1~+~\bf{s}\cdot\bf{h} 
\end{equation}
\noindent
where $\bf{s}$ is the tau polarization vector, and $\bf{h}$~is the polarimetric vector constructed using hadronic currents.
In our  $W^\pm$ and $H^\pm$ decays the
exact expression reduces to:

\noindent
\begin{equation}\label{eq:4}
w_\mathrm{T}=1~+~sign~h_z
\end{equation}
\noindent
where $sign$ equals one for left-handed taus from \CHW~ bosons and
minus one for right-handed taus from the charged Higgs boson.
$h_z$ is the $z$ component of the polarimetric vector.

In the \TauTau~final state, the weight is defined as~\cite{TAUOLA1,TAUOLA2,TAUOLA3}:

\noindent
\begin{equation}\label{eq:5}
w_\mathrm{T}=R_{\mathrm{ij}}h^{\mathrm{i}}h^{\mathrm{j}}
\end{equation}
\noindent
where $R_{\mathrm{ij}}$ is a matrix describing the full spin correlation
between the two taus as well as the individual spin states of the taus. $h^{\mathrm{i}}$ and $h^{\mathrm{j}}$
are the time ($t$) and space ($x,~y,~z$) components of the two taus' polarimetric vectors. The $t$ component of $h$ and $R_\mathrm{tt}$ are by convention set to $1$.

Neglecting the transverse spin degree, and in the ultrarelativistic limit, for \Ztautau~events the expression reduces to:

\noindent
\begin{equation}\label{eq:6}
w_\mathrm{T}=1~+~sign~h_{z^{+}}h_{z^{-}} +P_\tau h_{z^{+}} +P_\tau h_{z^{-}}
\end{equation}
\noindent
where \PTAU~denotes the polarization of the single tau 
in a mixed quantum state. Within this approximation,
\PTAU~is a linear function of the probability
\PTAUZ:
\noindent
\begin{equation}\label{eq:7a}
P_\tau=2\PTAUZ-1. 
\end{equation}
\noindent
In an event of a neutral and spin zero Higgs boson decaying to \TauTau, 
the expression simplifies to:

\noindent
\begin{equation}\label{eq:7}
w_\mathrm{T}=1~+~sign~h_{z^{+}}h_{z^{-}}
\end{equation}
\noindent
The $sign$ equals one for the \ZGAMMA~boson and minus one for the neutral Higgs boson,
reflecting the opposite spin correlations in the two samples.

Each tau decay channel requires a distinct method to calculate
the polarimetric vector~\cite{Davidson:2010rw}.
The tau decay modes implemented in {\tt TauSpinner} are 
listed in Tab.~\ref{tab:TauModes}. For the remaining channels, involving 
five pions in the final state and multi-prong decays with kaons, the
effect of tau polarization is neglected and $h_{z^{\pm}}$ is set to zero.
\begin{table}[h]
 \begin{center}
 \caption{Summary of tau decay modes implemented in the {\tt
   TauSpinner}. Branching fraction is given for each decay mode~\cite{PDG}.}
 \label{tab:TauModes}
  \begin{tabular}{lc}
\hline\noalign{\smallskip}
Tau decay mode & Branching fraction \%\\
\hline\noalign{\smallskip}
$e^-\bar{\nu_e}\nu_\tau$ & 17.85\\
$\mu^-\bar{\nu_\mu}\nu_\tau$ & 17.36\\
$\pi^-\nu$ &  10.91\\
$\pi^-\pi^0\nu$ &  25.51\\
$\pi^-\pi^0\pi^0\nu$, $\pi^-\pi^+\pi^-\nu$ &  9.29, 9.03 (incl. $\omega$)\\
$K^-\nu$ &  0.70\\
$K^-\pi^0\nu$, $\pi^- K^0\nu$ &  0.43, 0.84\\
$\pi^-\pi^+\pi^-\pi^0\nu$ &  4.54 (incl. $\omega$)\\
$\pi^-\pi^0\pi^0\pi^0\nu$ &  1.04\\
Other &  2.5\\
\hline\noalign{\smallskip}
  \end{tabular}
 \end{center}
\end{table}

\subsection{Application of the spin weight}
The event weight can be used at the analysis level for:
\begin{description}
\item[Simulating tau spin effects] e.g. in a sample generated
 without spin effects. The event weight equals $w_\mathrm{T}$. 
 It takes values between (0,2), except for the case of \Ztautau
 when the range is (0,4).
\item[Removing tau spin effects] from a sample generated
 with spin effects. The event weight equals 1$/w_\mathrm{T}$. Then, it 
 is greater than zero with no upper limit.
\item[Reverting tau spin effects] in a sample generated
 with certain longitudinal tau polarization (and/or correlations) to
 the different one. The weight equals
 (2-$w_\mathrm{T}$)/$w_\mathrm{T}$ for the \TauNu~final state 
 corresponding to the \CHW$\to$\CHHIGGS~or \CHHIGGS$\to$\CHW~replacement,
 and  $w_\mathrm{T}$(\HIGGS)/$w_\mathrm{T}$(\ZGAMMA)
 or $w_\mathrm{T}$(\ZGAMMA)/$w_\mathrm{T}$(\HIGGS)
 for the \TauTau\\final state corresponding to \ZGAMMA$\to$\HIGGS~or \HIGGS$\to$\ZGAMMA~
  replacement. It is greater than zero without an  upper limit.
\end{description}

\subsection{Attributing tau helicity states in the \Ztautau decays}

The helicity states of the taus are attributed stochastically
by comparing a random number with the probability, \PTAUZ~$w_\mathrm{T}($\PTAUZ$=1)/w_\mathrm{T}$, of the right-handed configuration to occur. 
This method is valid for events generated
with spin effects or emulated using the spin weight provided by the {\tt TauSpinner}. An average over all possible initial state configuration is taken.
Presentation  of the method principle can be found in Ref.~\cite{Davidson:2010rw}.
\section{Performance of the {\tt TauSpinner} algorithm}\label{sect:Perf}

The performance of the {\tt TauSpinner} algorithm is studied based on the
MC simulated \Ztautau~and \Wtaunu~events.
The samples were generated using the general purpose event generator
{\tt Pythia}~\cite{PYTHIA} assuming the $pp$ collision at center of mass energy of 7 TeV.
The taus were then made to decay using the {\tt Tauola} package. Two sets of events were simulated:
\begin{description}
\item[{\it No spin effects.}] In these events the taus were decayed by {\tt Tauola} as if they had been produced with no
polarization. They were used for emulation of spin effects later, with the {\tt TauSpinner} package. 
\item[{\it Tauola.}] In these events tau spin effects were properly accounted for at the time of 
tau event generation and decay.
\end{description}

Event generation and simulation of the spin effects in the {\tt TauSpinner} were performed 
with the modified LO parton distribution function (PDF)
MRSTLO*~\cite{PDF}.

\subsection{Simulation of the tau polarization in \TauNu~final state}

For all tau decay modes, the main observables that are sensitive to 
the tau polarization are the tau momentum fraction taken by the
hadronic system~\cite{B235}, x, and the relative difference between the 
charge and the neutral energy in the tau decay, $\Upsilon$~\cite{POLW}.

\begin{figure}
\centering
 \subfigure[Fraction of the tau momentum taken by the hadron in the combined \PI~and \KA~channels.]{\includegraphics[width=0.45\textwidth]{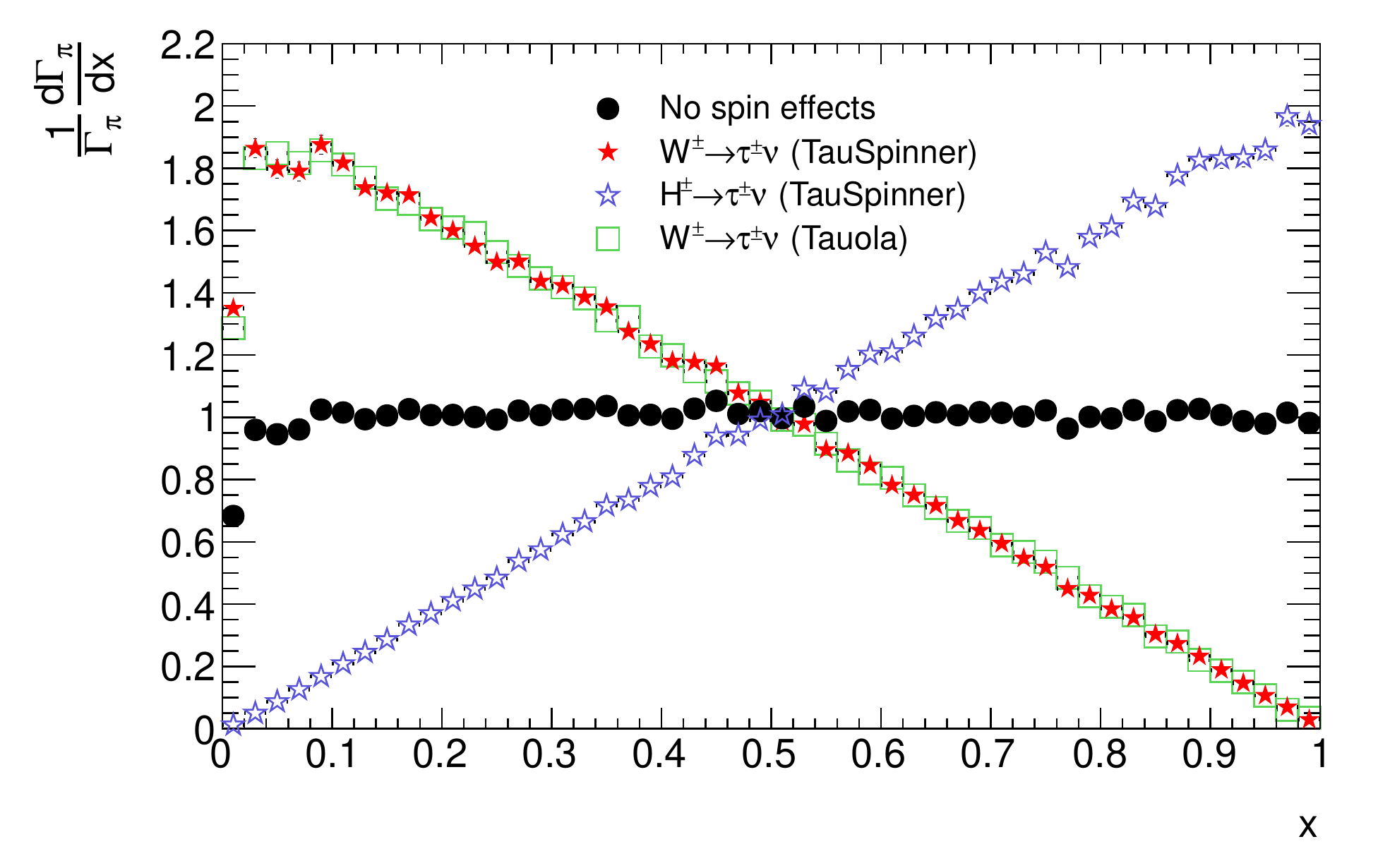}\label{fig:TauPolar1}}
 \subfigure[Relative difference between the charged and the neutral energy in the \RHO~channel.]{\includegraphics[width= 0.45\textwidth]{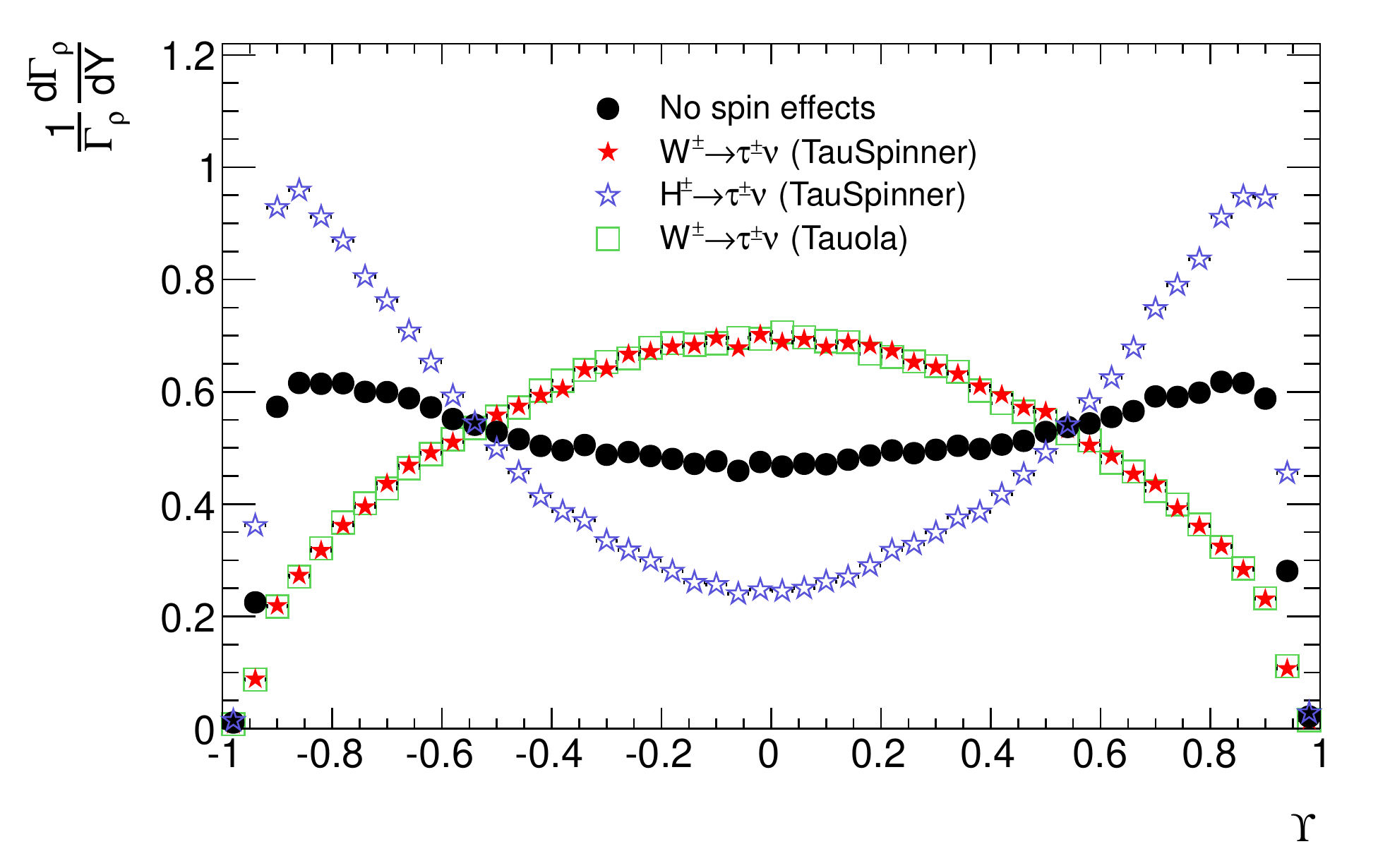}\label{fig:TauPolar2}}

 \subfigure[Relative difference between the charged and the neutral energy in the combined \ARES~and \ARESMULTI~channels.]{\includegraphics[width=0.45\textwidth]{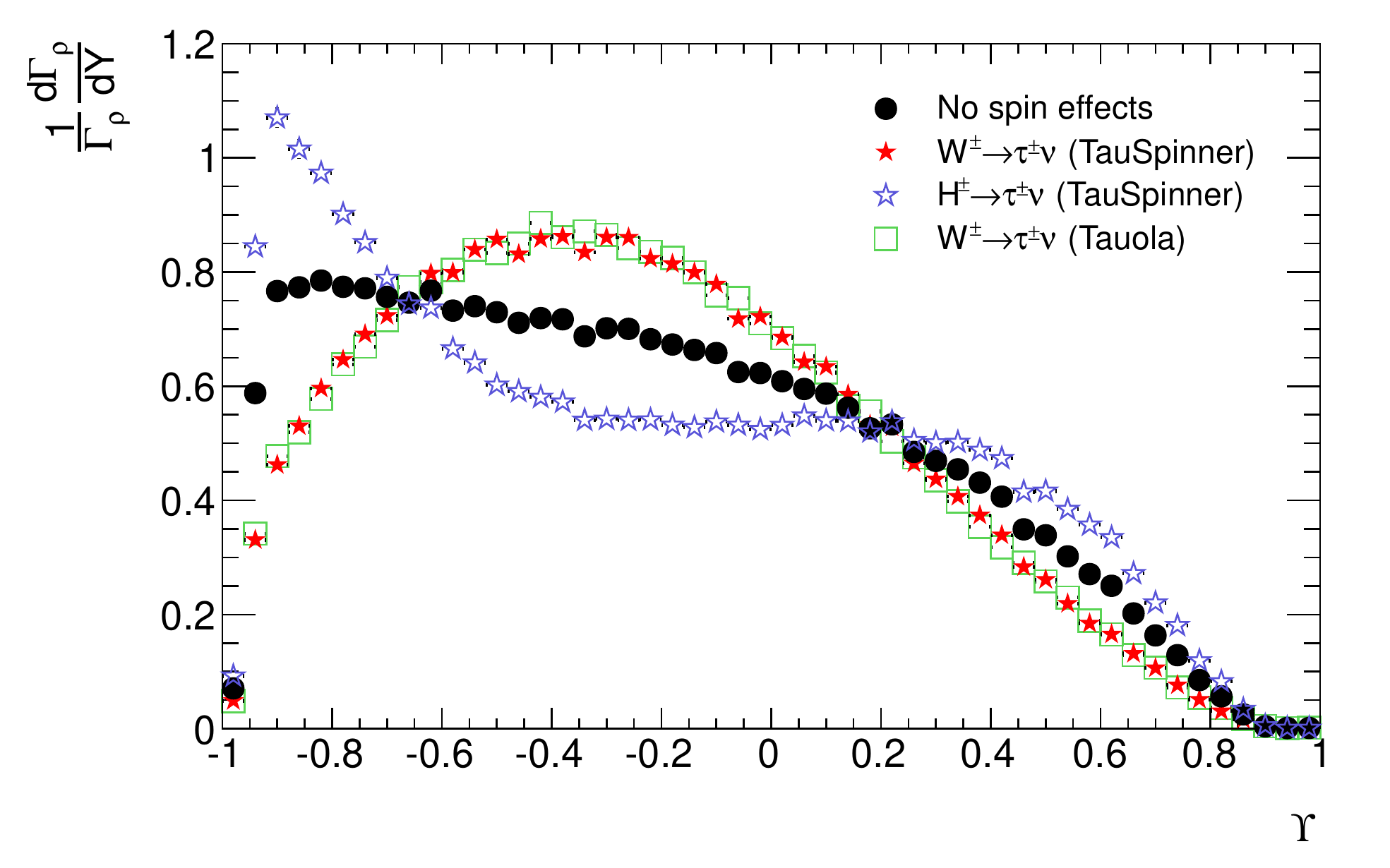}\label{fig:TauPolar3}}
\subfigure[Relative difference between the
  charged and the neutral energy in the \RHO~channel. Decays of taus in the ``no spin effects'' sample were generated using {\tt Pythia} 6.425.]{\includegraphics [width=0.45\textwidth]{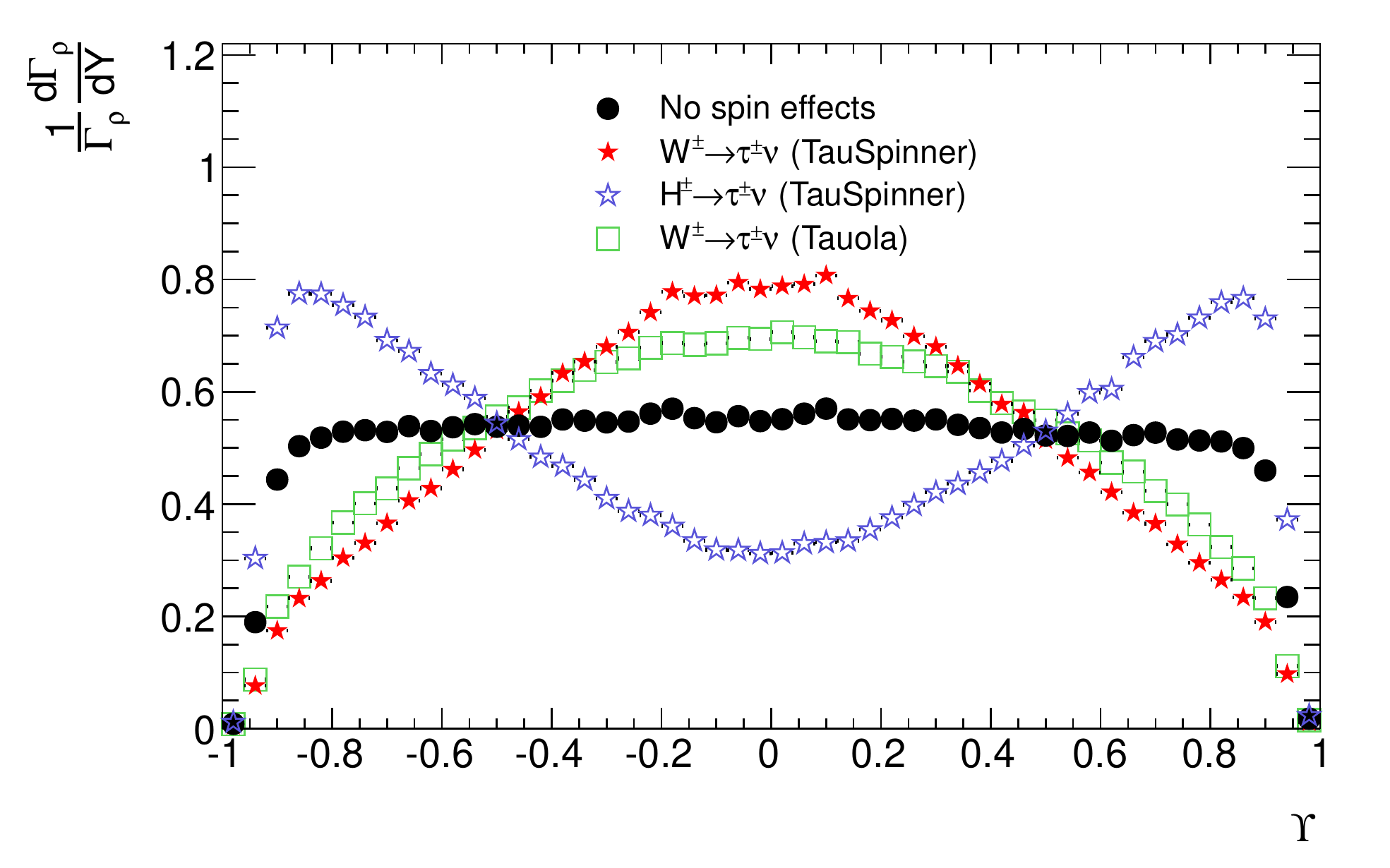}}
 \caption{Comparison of spin effects simulated with {\tt Tauola} and {\tt TauSpinner} for various observables.}
 \label{fig:TauPolar}
\end{figure}

Plots a)--c) in Figure~\ref{fig:TauPolar} demonstrates the performance of  {\tt TauSpinner}
for the channels where the tau polarization is extremal.
The observable x is plotted in Fig.~\ref{fig:TauPolar1} 
for the combined \PI~and \KA~channels. The observable $\Upsilon$~is plotted in Fig. 
~\ref{fig:TauPolar2} and ~\ref{fig:TauPolar3} for the \RHO~and
combined \ARES~and \ARESMULTI~channels, respectively.
The sample with no spin effects refers to \Wtaunu\\events generated
with flat \PTAU~value. The \Htaunu~and \Wtaunu~configurations were obtained by applying an
appropriate spin weight to the sample with no spin effects.
The weighted observables exhibit the expected behavior
indicating a proper implementation of the \PTAU~in the {\tt TauSpinner} package.

Note that the {\tt TauSpinner} algorithm requires that the decays of unpolarized taus 
via $\rho$~or $a_1$~preserve the spin correlation between the production and decay of 
these mesons. This feature is missing if the taus are made to decay using
{\tt Pythia} 6.425. Although a bulk of spin effects are reconstructed, a significant 
systematic uncertainty arises as demonstrated in
Fig.~\ref{fig:TauPolar} d).

\subsection{Simulation of tau polarization in \Ztautau events}

\begin{figure}
\centering
 \subfigure[The up quark aligned along the positive $z$ axis.]{\includegraphics[width=0.45\textwidth]{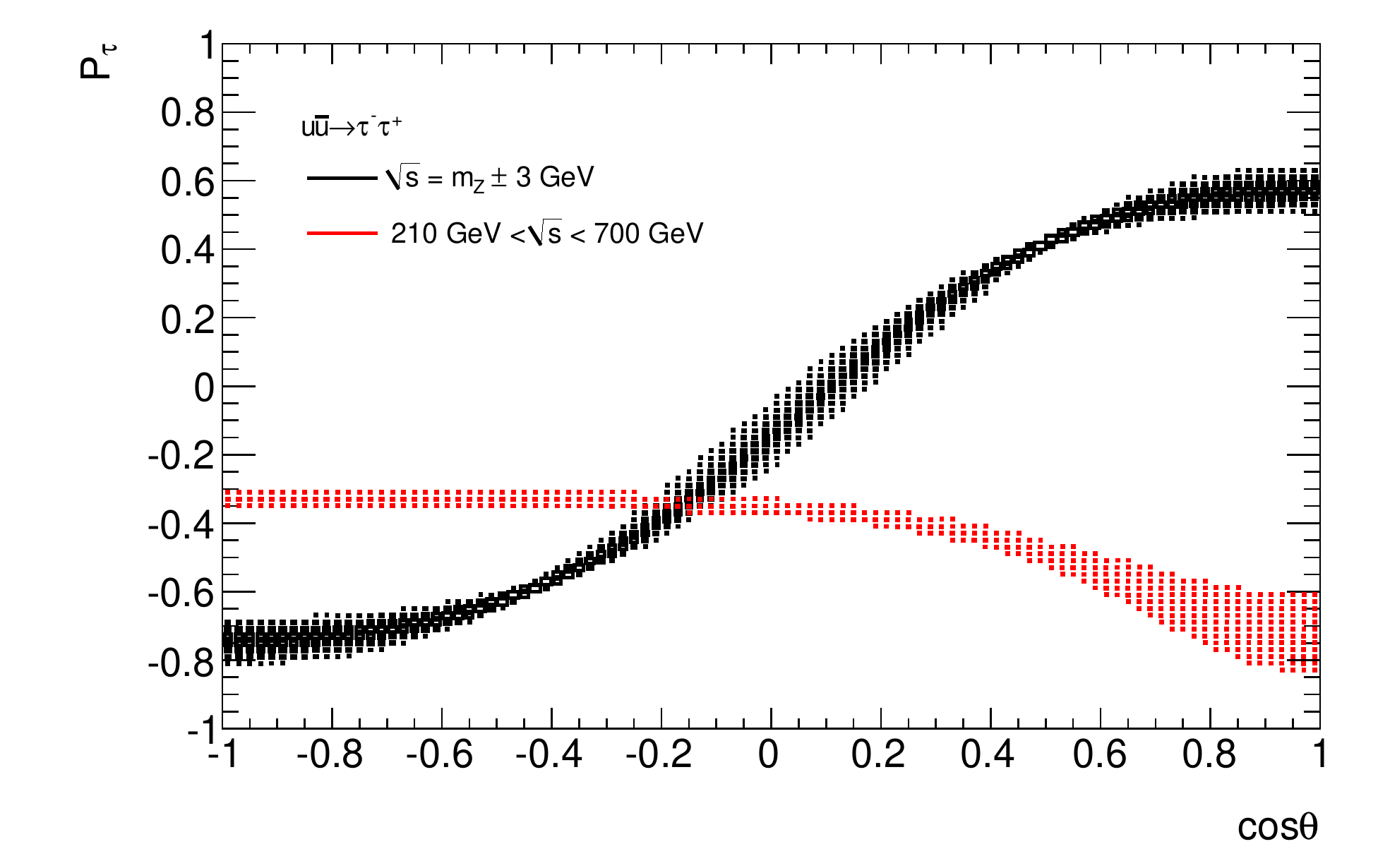}\label{fig:TauPolCos1}}
 \subfigure[The down quark aligned along the positive $z$ axis.]{ \includegraphics[width=0.45\textwidth]{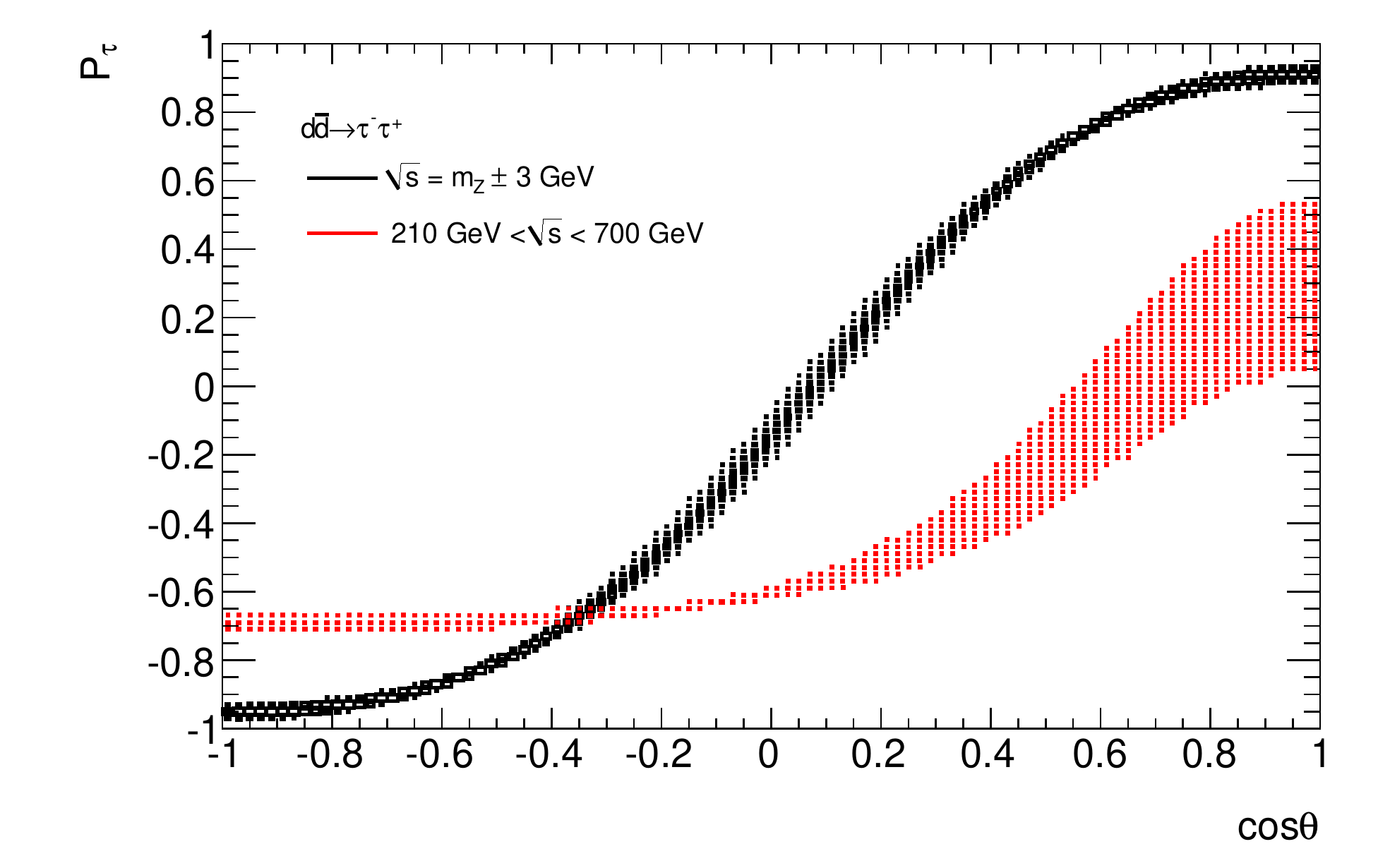}\label{fig:TauPolCos2}}
 \caption{Tau polarization as a function of $\cos\theta$.}
 \label{fig:TauPolCos}
\end{figure}

In \Ztautau~events, the single tau polarization \PTAU~depends on the intermediate boson virtuality,
the flavor of the incoming quark and the scattering angle.
The complexity of this dependence is shown in
Fig.~\ref{fig:TauPolCos} where the tau polarization is drawn as a function 
of $\cos\theta$, for two different ranges of the invariant mass of the boson.
In these plots, the quark flavor configuration is fixed at the level of simulation of the spin weight in the {\tt TauSpinner}.

\begin{figure}[h]
\centering
\subfigure[The up and down quarks aligned along the positive z axis.]{\includegraphics[width=0.45\textwidth]{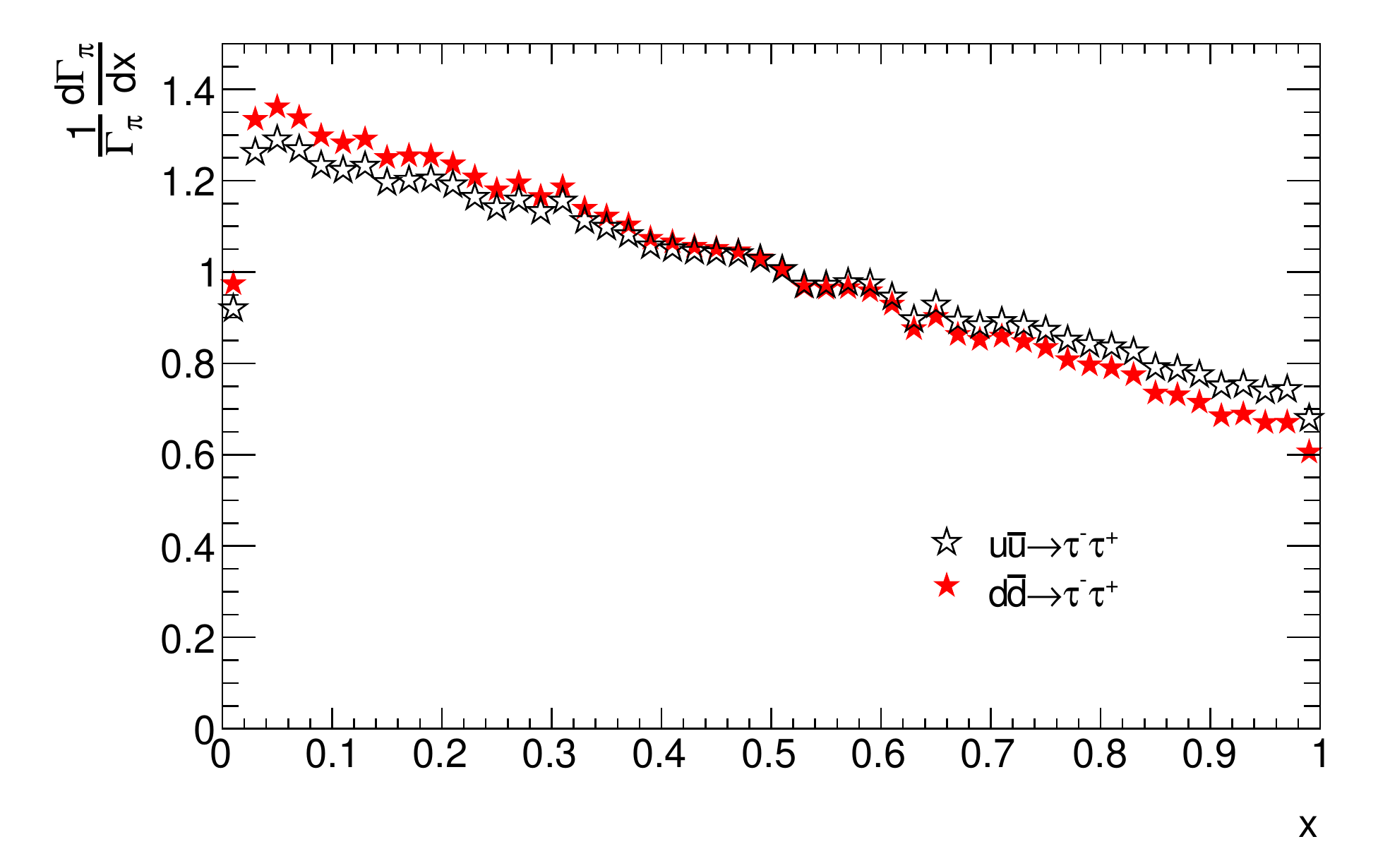}\label{fig:spinA}}
\subfigure[The up and down quarks aligned along the negative z axis.]{\includegraphics[width=0.45\textwidth]{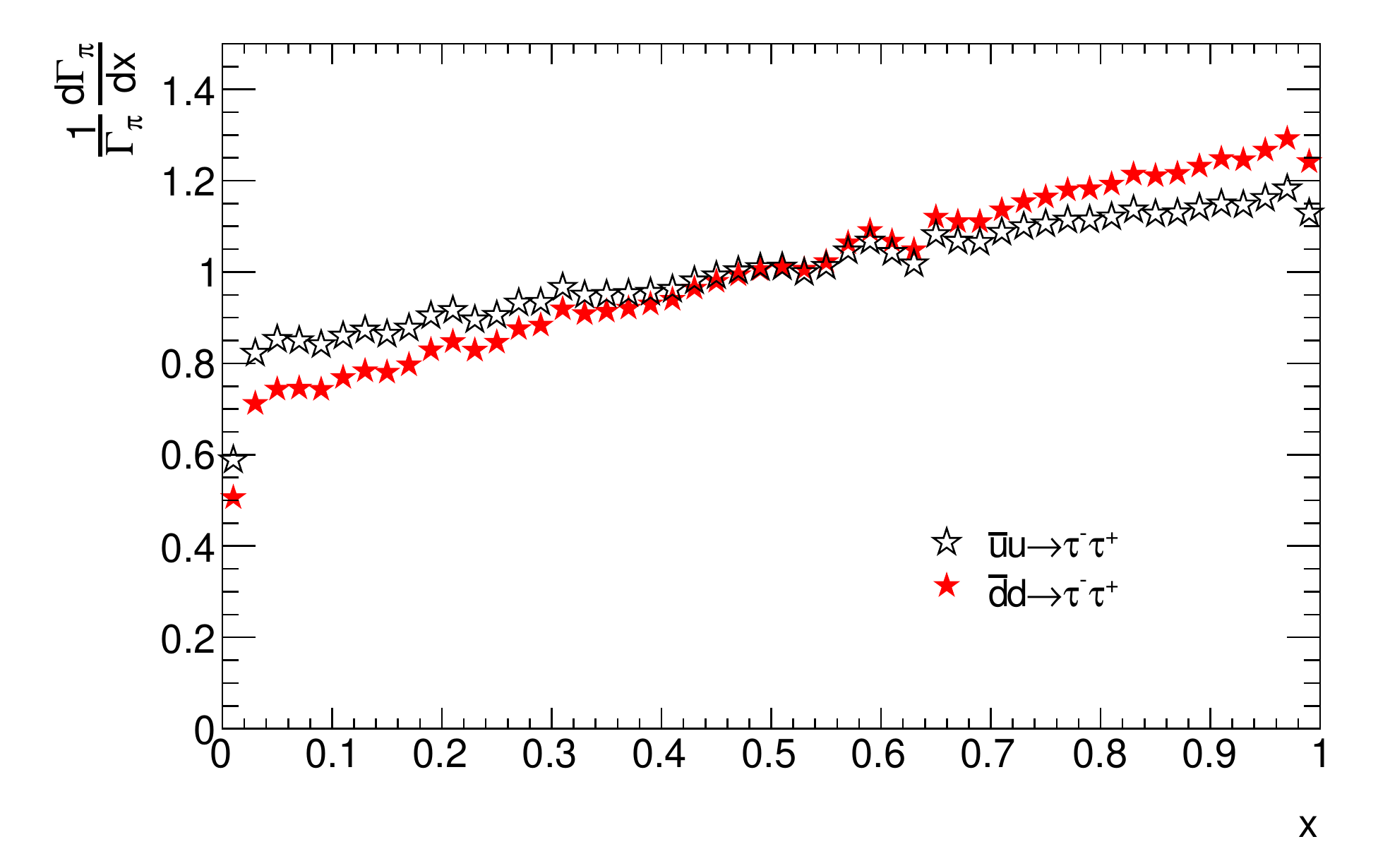}\label{fig:spinB}}
\caption{Fraction of the tau momentum taken by the hadron in the \PI~and \KA~channel in \Ztautau~events. Taus with negative charge emitted in the forward hemisphere is chosen.}
\label{fig:XfrICC}
\end{figure}
\begin{figure}
\centering
\subfigure[$Z$ boson emitted in the forward direction.]{\includegraphics[width=0.45\textwidth]{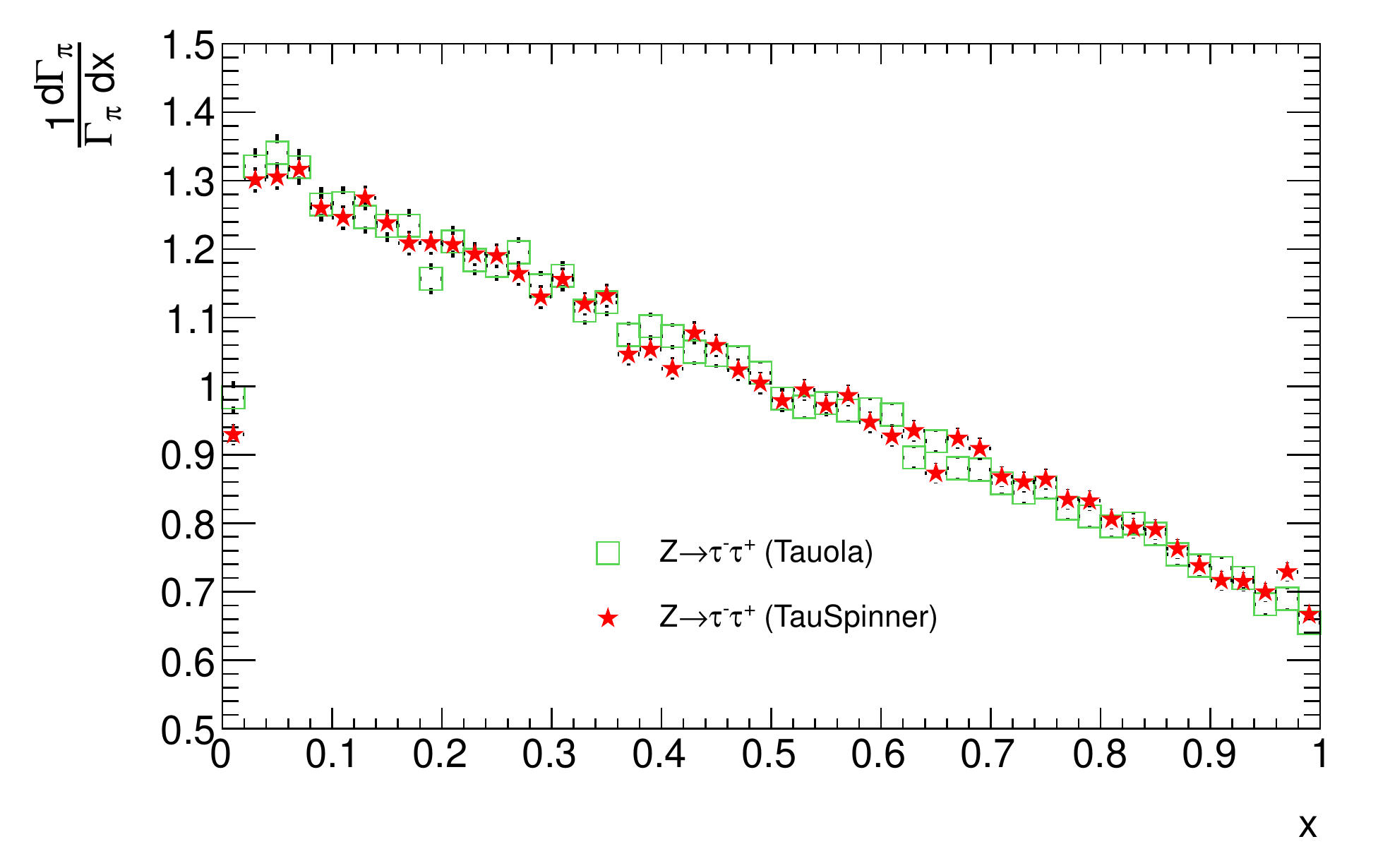}\label{fig:spinA1}} 
\subfigure[$Z$ boson emitted in the backward direction.]{\includegraphics[width=0.45\textwidth]{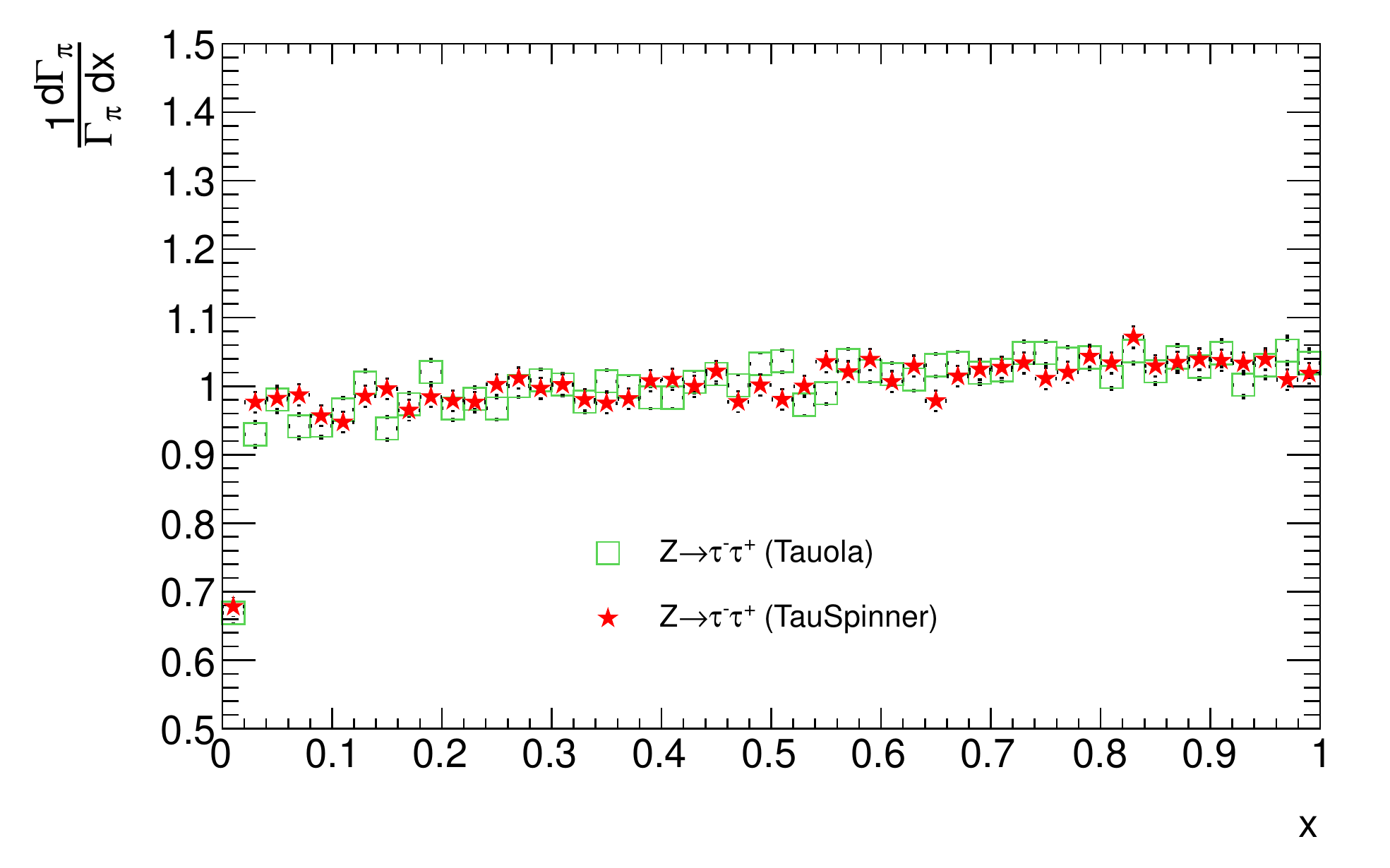}\label{fig:spinB1}}
\caption{
Fraction of the tau momentum taken by the hadron in the \PI~ 
channel. Taus with negative charge emitted in the forward hemisphere are chosen.
}
\label{fig:XfrFB}
\end{figure}

In the next step, a degree of polarization is studied for different configurations
of the incoming quarks. The virtuality of the intermediate boson is constrained to lie within $\pm$3 GeV
of the $Z$ boson mass
and both taus are set to decay to \PI~or \KA.
The forward-backward spin asymmetry is accessed by choosing the
\TauMin~to be emitted in the forward region by requiring
the longitudinal momentum of \TauMin~to be greater than that of \TauPlus. 
Figure~\ref{fig:XfrICC} shows the observable $x$ for the up and down type quarks
entering along the positive or negative $z$ axis.
The results are consistent with those of reference~\cite{TauSpinERWZW},
and therefore reassure a proper transmission of spin effects from the hard process
to the tau decay products.

In the last step, the degree of polarization is
studied inclusively for all initial state quark configurations and the results are compared to
those simulated using the {\tt Tauola} package. The \ZGAMMA~virtuality is constrained to lie within $\pm$3 GeV of the $Z$ boson mass.
The fraction of the tau momentum taken by the hadron in the \PI~channel is plotted in Fig.~\ref{fig:XfrFB}. Expecting the $Z$ boson to 
be emitted in the direction of the incoming quark (as compared to the
direction of the anti-quark from the sea) these plots can be compared to those in Fig.~\ref{fig:XfrICC}.

\begin{figure}
 \centering
  \includegraphics[width=0.45\textwidth]{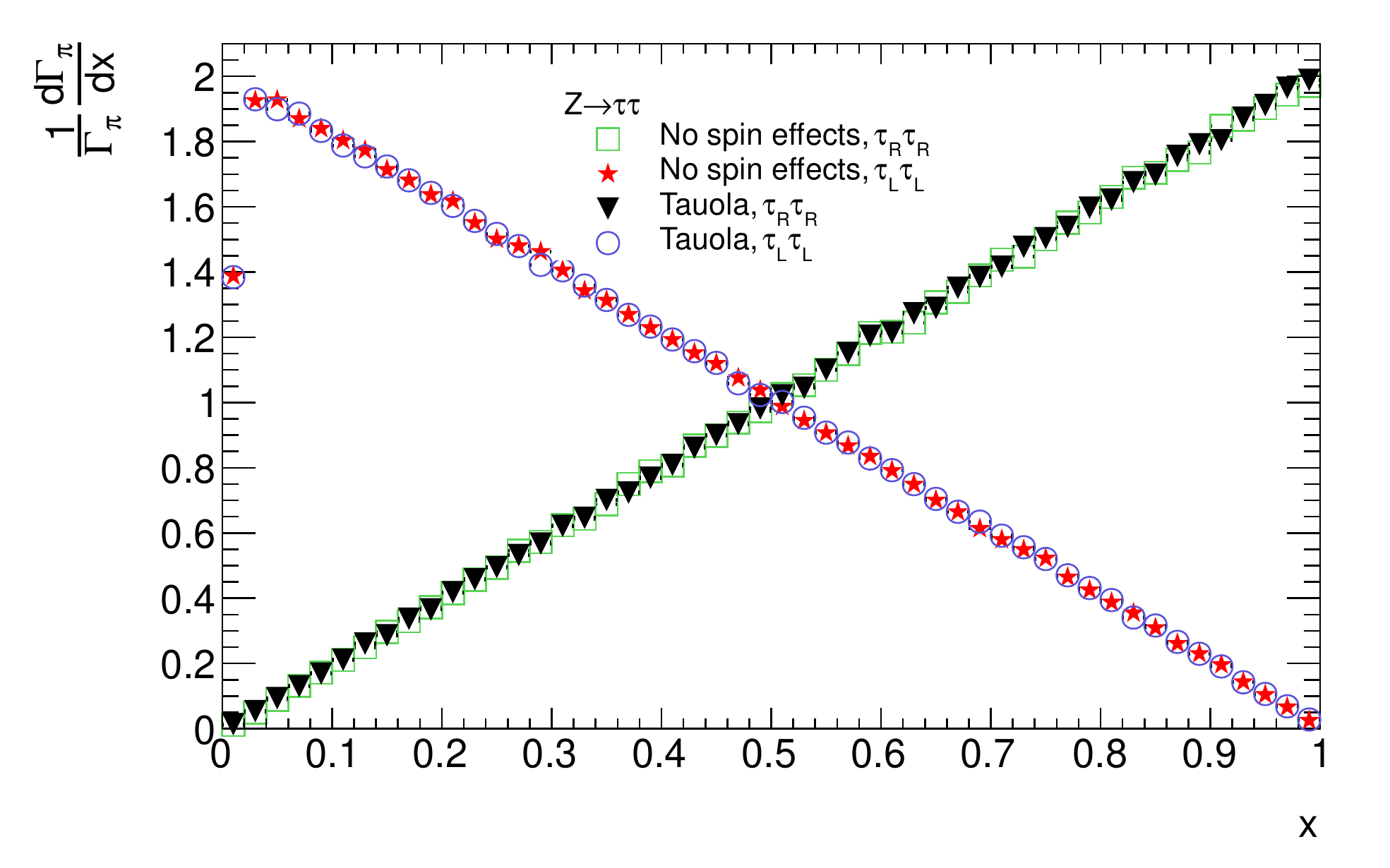}
  \caption{Fraction of the tau momentum taken by the hadron in the combined \PI~and \KA~channels. $\tau_R$ and $\tau_L$ denote right-handed and left-handed taus, respectively.}
\label{fig:TauSplitting}
\end{figure}

Figure~\ref{fig:TauSplitting} shows the fraction of the tau momentum taken by the hadron in the combined \PI~and \KA~channels
for the left-handed and the right-handed taus from the \Ztautau decays. Helicity states are attributed by the {\tt TauSpinner}
in the sample without and with spin effects simulated at the generation level.
The distributions exhibit the expected shapes for properly attributed pure helicity states.
\subsection{Validation of spin correlation in the \TauTau final state}

\begin{figure}
\centering
 \subfigure[ The z$_s$ variable described in text in the channel where both taus decay to \PI~or \KA.]{\includegraphics[width=0.45\textwidth]{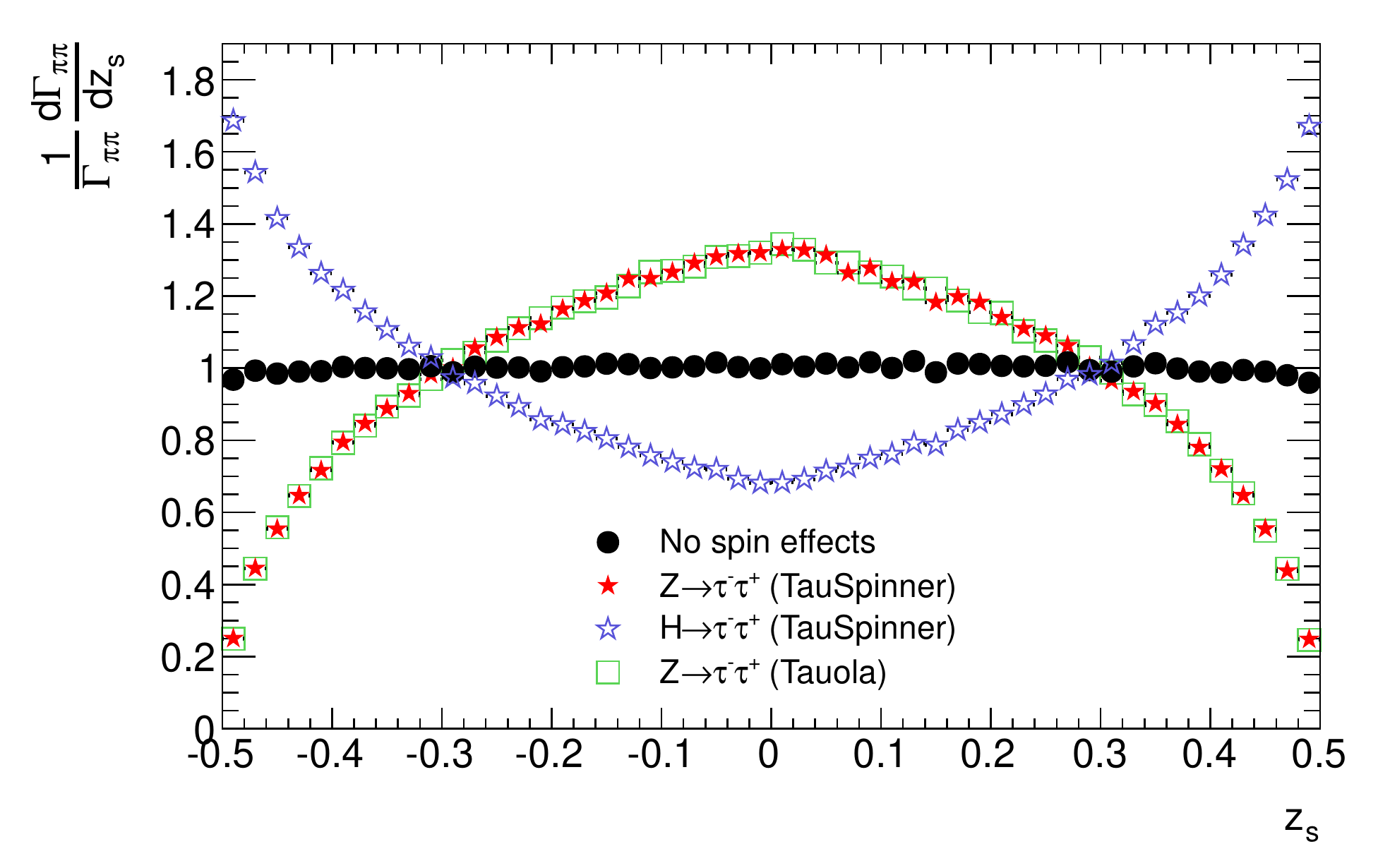}\label{fig:TauCorr1}}
 \subfigure[ Visible mass of the two hadrons in the channel where both taus decay to \PI~or \KA.]{\includegraphics[width=0.45\textwidth]{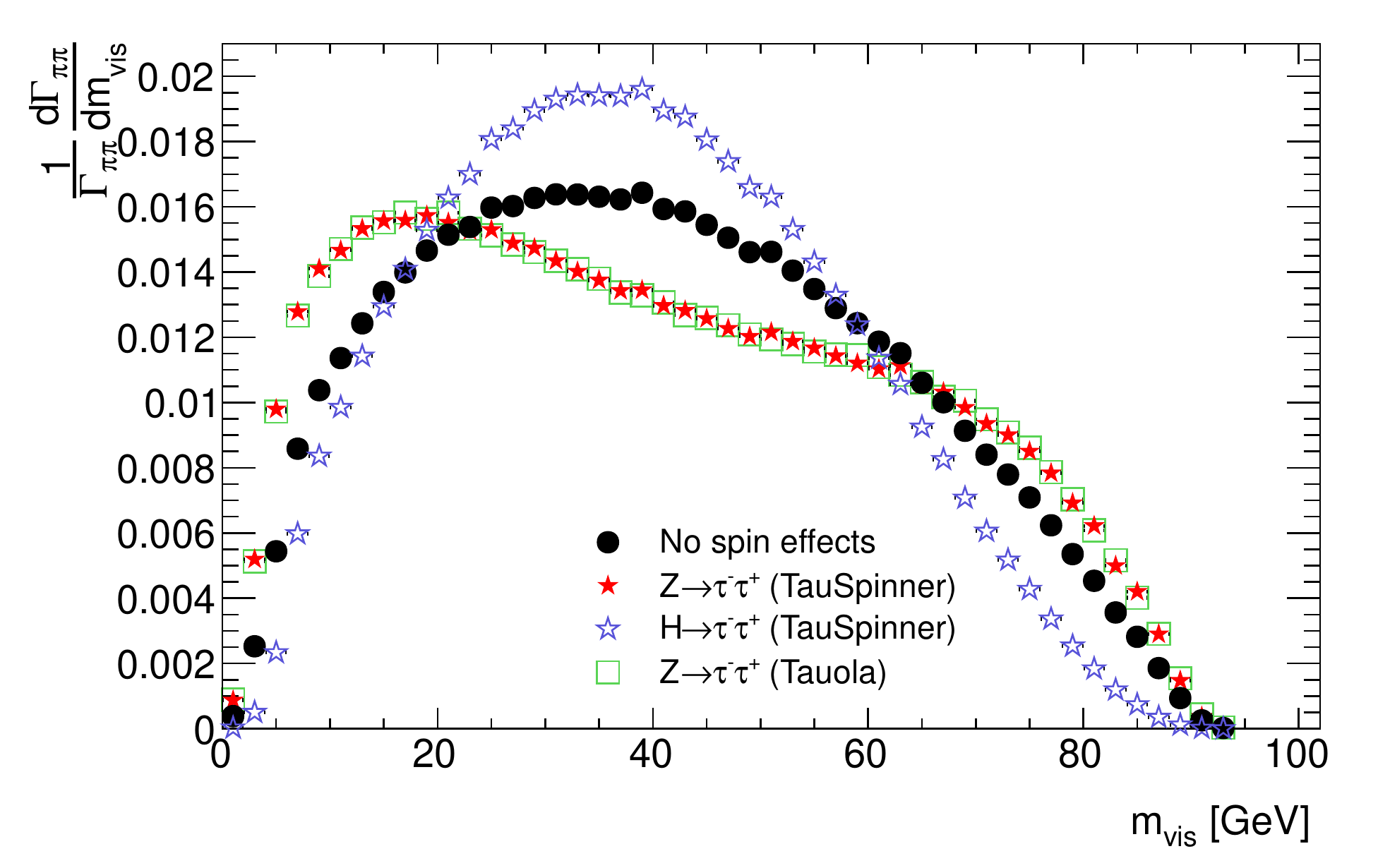}
  \label{fig:TauCorr2}
 }
 \caption{Tau spin correlation observables.}
 \label{fig:TauCorr}
\end{figure}

Two observables were demonstrated in~\cite{TauSpinERWZW} to be appropriate for studying 
tau spin correlations in the \TauTau~final state: the
invariant mass of the hadronic system, $m_\mathrm{vis}$, and the z$_s$ variable.
The latter is a signed surface in the x$^+$~- x$^-$ plane, between 
lines: x$^+$=x$^-$ and x$^+$=x$^-$+a. x$^+$ and x$^-$ 
denote the fraction of tau momenta taken by the hadrons.
The two observables are plotted in Fig.~\ref{fig:TauCorr}
in the channel where both taus decay to \PI~or \KA.
The sample with no spin effects refers to \Ztautau~events generated
with \PTAUZ=0.5~and no spin correlation. The proper \Htautau~and \Ztautau~configurations were obtained by applying an
appropriate spin weight to this sample without spin effects.
The weighted observables exhibit the expected behavior
reassuring a proper implementation of the spin correlations in the {\tt TauSpinner} package.

\section{Summary and outlook}\label{sect:Summary}

The {\tt TauSpinner} package designed to emulate tau spin effects has
been introduced. The algorithm is limited to the leading order accuracy and
the longitudinal spin degrees only. As compared to the algorithm developed in Ref.~\cite{TauSpinERWZW}, the
functionality is extended to allow one to estimate the spin effects when the
information on the incoming quarks entering the hard process is not
available. This is therefore the first algorithm suitable for emulating the spin effects
in the tau-embedded samples. The intrinsic tau polarization arising from parity violation in the weak interactions
is attributed on the basis of intermediate boson kinematics and PDF's.
For the \Ztautau~process its complete functionality is limited to $pp$ collisions
and otherwise restricted to conservation of total angular momentum. 

Comparisons of {\tt Tauola} and {\tt TauSpinner} for various spin observables
presented in Sec.~\ref{sect:Perf} demonstrate 
that no additional systematic uncertainty in simulation of 
tau spin effects has been introduced in the {\tt TauSpinner}. 
This claim should, however, be validated in a sample
where the number of high transverse momentum jets is enhanced.

A complete discussion of the theoretical uncertainties
is common for the {\tt Tauola}, the {\tt TauSpinner} and, to a large extent
the KORALZ~\cite{Jadach:1993yv} MC programs.
Some aspects of these uncertainties have been addressed through
the work published in Refs.~\cite{TAUOLA1,TAUOLA2,TAUOLA3,Jadach:1984iy,Jadach:1999vf,vanHameren:2008dy}.
A rigorous evaluation of theoretical effects and in particular a comparison
with results based on the next-to-leading order matrix element calculations
simultaneous for scattering processes and
for spin density matrices is referred to a future work.


Although the current version of the {\tt TauSpinner} is restricted to
the longitudinal spin degrees, the framework is prepared to
simulate the complete spin effects. This extension is planned for the future version of the algorithm.
The code of {\tt TauSpinner} is publicly available, 
with all relevant details given in App. A and B.

\section{Acknowledgements}
\label{sec:Ack}
We would like to thank A. Buckley, S.Demers, C. Gwenlan, E. Richter-Was and S. Tsuno for valuable discussions. 



\begin{thebibliography}{10}

\bibitem{EMBEDDED}
{ATLAS Collaboration}, Phys. Lett. B \textbf{705}, 174 (2011).

\bibitem{TAUOLA1}
{S. Jadach et al.}, Comput. Phys. Commun. \textbf{64}, 275 (1990).

\bibitem{TAUOLA2}
M.~Jezabek, et~al., Comput. Phys. Commun. \textbf{70}, 69 (1992).

\bibitem{TAUOLA3}
R.~Decker, et~al., Comput. Phys. Commun. \textbf{76}, 361 (1993).

\bibitem{TauSpinERWZW}
T.P. et~al., Acta Phys. Pol. B \textbf{32}, 1277 (2001).

\bibitem{Davidson:2010rw}
N.D. et~al., arXiv:1002.0543  (2010).

\bibitem{Jadach:1984iy}
{S. Jadach and Z. Was}, Comput.Phys.Commun. \textbf{36}, 191 (1985).

\bibitem{Jadach:1993yv}
{S. Jadach, B. Ward and Z. Was}, Comput.Phys.Commun. \textbf{79}, 503 (1994).

\bibitem{Jadach:1999vf}
{S. Jadach, B. Ward and Z. Was}, Comput.Phys.Commun. \textbf{130}, 260 (2000).

\bibitem{Was:1989ce}
{Z. Was and S. Jadach}, Phys. Rev. D \textbf{41}, 1425 (1990).

\bibitem{PDG}
{K. Nakamura et al.}, J. Phys. G \textbf{37}, 260 (2010).

\bibitem{PYTHIA}
{T. Sjostrand}, Comput. Phys. Commun. \textbf{82}, 74 (1994).

\bibitem{PDF}
{A. Sherstnev, R.S. Thorne}, Eur. Phys. J. C Part. Fields \textbf{55}, 553
  (2008).

\bibitem{B235}
{K. Hagiwara, A.D. Martin, D. Zeppenfeld}, Phys. Lett. B \textbf{235}, 198
  (1990).

\bibitem{POLW}
{ATLAS Collaboration}, ATLAS-CONF-2012-009  (2012).

\bibitem{vanHameren:2008dy}
{A. van Hameren and Z. Was}, Eur. Phys. J. C \textbf{61}, 33 (2009).

\bibitem{TauSpinnerOfficial2}
{Z. Czyczula, T. Przedzinski, Z. Was },\\
  http://wasm.web.cern.ch/wasm/Welcome.html\\
  http://hibiscus.if.uj.edu.pl/$\sim$przedzinski/tau-reweight

\bibitem{LHAPDF}
{M.R. Whalley, D. Bourilkov and R.C. Group}, arXiv:hep-ph/0508110  (2005).

\bibitem{Dobbs:2001ck}
{M. Dobbs and J.B. Hansen}, Comput. Phys. Commun. \textbf{134}, 41 (2001).

\end{thebibliography}
\newpage
\clearpage
\onecolumn

\appendix
\section{Requirements for data files}
\label{App:information}

Data files generated by a MC event generator (or constructed using the tau-embedding method) 
need to fulfill the following requirements: 
\begin{enumerate}
\item Four-momenta of the intermediate boson, the taus and the flavor and the four-momenta of the tau decay products need to be available.
\item Flavor of the intermediate boson needs to be available or set by the user.
\item For all types of hard processes, but \Ztautau, the four-momenta can be defined
in an arbitrary but common frame. For the \Ztautau~process, the four-momenta have to be given
in the laboratory frame in order to be consistent with the PDFs.
\item The four-momenta of the taus and their decay products need to be known with sufficient precision in order to
ensure numerical stability of the algorithm. Six significant digits are recommended.
\end{enumerate}

Note that different MC generators may store the 
truth information in different ways. It is the responsibility of a user to make sure
that all these requirements are fulfilled.

\section{Public version}

A generic version of the package can be found in~\cite{TauSpinnerOfficial2}.
The main code is written in C++ and relies upon two libraries:
{\tt Tauola} and {\tt LHAPDF}~\cite{LHAPDF}. A method for reading input information stored using the {\tt HepMC}~\cite{Dobbs:2001ck} format is prepared.
Support for any other input format is available upon request.

\subsection{Project organization}
\addcontentsline{toc}{part}{Appendices}

The {\tt TauSpinner} package is organized in the following manner:
\begin{itemize}
\item {\tt src/tau\_reweight\_lib.c, src/tau\_reweight\_lib.h} - the core of the algorithm.
\item {\tt src/Tauola\_wrapper.h} - wrapper for {\tt TAUOLA FORTRAN} routines.
\item {\tt src/SimpleParticle.h} - definition of {\tt {\bf class} SimpleParticle} used
   as a bridge between the event record (or data file) and the algorithm.
\item {\tt src/Particle.h} - definition of {\tt {\bf class} Particle} used
   for boosting and rotation of the particles.
\item {\tt src/read\_particles\_from\_TAUOLA.c, src/read\_particles\_from\_TAUOLA.h} - interface to \\
   the {\tt HepMC::IO\_GenEvent} data files used by the example program.
\item {\tt README} - a short manual.
\end{itemize}

\subsection{The algorithm sequence}\addcontentsline{toc}{part}{Appendices}

The {\tt TauSpinner} takes the following sequence of steps:
\begin{description}
\item [Initialization of {\tt Tauola}.] It is ensured by invoking:\\ 
   {\tt Tauola::initialize(); }
\item [Initialization of {\tt TauSpinner}.] It is performed by executing:\\
   {\tt void initialize\_spinner({\bf bool} Ipp, {\bf int} Ipol, {\bf double} CMSENE)} \\
   where the argument {\tt Ipp} passes the information on the type of collision events ({\tt Ipp = true} sets $pp$ collisions),
   {\tt Ipol} passes the information on the spin effects included in the input sample ({\tt Ipol}= 0, 1, 2 corresponds to no spin effects, complete spin effects
   and spin correlations only, respectively)  and
   {\tt CMSENE} sets the collision center of mass energy.
\item [Reading the data files.] Information on the four-momenta and the flavor of the boson,
  the final state taus or tau and neutrino pair and the tau decay products
  is filled and stored in instances of {\tt SimpleParticle} class by
  the use of the function:\\
  {\tt {\bf void} readParticlesFromTAUOLA\_HepMC(HepMC::IO\_GenEvent \&input\_file, SimpleParticle \&boson,\\ SimpleParticle \&tau, SimpleParticle \&tau2, vector$<$SimpleParticle$>$ \&tau\_daughters,\\ vector$<$SimpleParticle$>$ \&tau2\_daughters)}.\\
This function should be modified if the input files are not in the 
{\tt HepMC::IO\_GenEvent} format.
\item [Calculation of the spin weight.] It is performed by the use of the following functions:\\
   {\tt {\bf double} calculateWeightFromParticlesWorHpn(SimpleParticle \&boson, SimpleParticle \&tau,\\ SimpleParticle \&tau2, vector$<$SimpleParticle$>$ \&tau\_daughters)} for the \Wtaunu~and\\ \Htautau~processes\\
   {\tt {\bf double} calculateWeightFromParticlesH(SimpleParticle \&boson, SimpleParticle \&tau,\\ SimpleParticle \&tau2, vector$<$SimpleParticle$>$ \&tau\_daughters,\\ vector$<$SimpleParticle$>$ \&tau2\_daughters)} for the \Ztautau~and \Htautau~processes. 
\item [Attributing tau helicity states.] For \Ztautau process, the tau helicity states
are attributed at the stage of calculation of the spin weight.
The information can obtained by calling {\tt getTauSpin()} function.
\end{description}

\subsection{Calculation of the spin weight}

For the \TauNu~final states, the spin weight is calculated in the following steps:

\begin{enumerate}
\item The parent boson, the tau, the tau neutrino and the list of tau daughters are identified
and boosted to the \TauNu~rest frame in which the tau is aligned along the $z$ axis.
\item The tau daughters are boosted to the tau rest frame. Two angles of spacial orientation
of the neutrino from the tau decay, {\tt theta2} and {\tt phi2}, are calculated and stored.
The tau daughters are rotated by these angles in order to align the neutrino along the $z$ axis.
\item The {\tt Tauola} decay channel is identified.
\item The {\tt Tauola FORTRAN} subroutine is called to perform calculation of the polarimetric vector $h$.
\item The polarimetric vector $h$ is rotated back using the {\tt theta2} and {\tt phi2} angles.
\item The spin weight is calculated using eq.~\ref{eq:4} and returned to the main program.
\end{enumerate}

\noindent For the \TauTau~final states, the spin weight is calculated in the following steps:

\begin{enumerate}
\item The parent boson, the taus and their tau daughters are identified
and boosted to the \TauTau~rest frame in which the taus are aligned along the $z$ axis.
\item For each tau:
\subitem Its identified daughters are boosted to its rest frame. Two angles of spacial orientation
of the neutrino from the tau decay, {\tt theta2} and {\tt phi2}, are calculated and stored. 
The tau daughters are rotated by these angles to align the neutrino along the $z$ axis.
\subitem The {\tt Tauola} decay channel is identified.
\subitem The {\tt Tauola FORTRAN} subroutine is called to perform calculation of the polarimetric vector $h$.
\subitem The polarimetric vector $h$ is rotated back using the {\tt theta2} and {\tt phi2} angles
\item In case of the \Ztautau~decays:
\subitem The probability \PTAUZ~is calculated using eqs~\ref{eq:1}-\ref{eq:2}.
\subitem The spin weight is calculated using eq.~\ref{eq:6} and returned to the main program.
\item In case of the \Htautau~decays:
\subitem The spin weight is calculated using eq.~\ref{eq:7} and returned to the main program.
\end{enumerate}

\subsection{The LHAPDF library wrapper}
\addcontentsline{toc}{part}{Appendices}

The evolution of the PDF's is invoked from the wrapper for PDF's: \\
\\
{\tt {\bf double} f(double x, int ID, double SS, double cmsene) } \\
\\
\noindent where function {\tt f} calls the evolution function {\tt xfx(x, SS, ID)}~\cite{LHAPDF}.
The PDF sets need to be available locally. They can be obtained from the {\tt LHAPDF} project website.

\end{document}